\documentstyle[11pt,aaspp4]{article}

\def\mj{M$_{Jup} $}

\def\deg{^{\circ}}
\def\cd{CoD $-$33$\deg$7795 }

\def\ha{H$\alpha$ }

\begin{document}
 
\title{A Candidate Substellar Companion to CoD $-$33$\deg$7795 (TWA 5)}

\author{Patrick J. Lowrance$^1$, Chris McCarthy$^1$, E. E. Becklin$^1$, 
B. Zuckerman$^1$, Glenn Schneider$^2$, R. A. Webb$^1$,
Dean C. Hines$^2$, J. Davy Kirkpatrick$^3$, David W. Koerner$^4$, Frank Low$^2$, 
Roland Meier$^5$,Marcia Rieke$^2$,Bradford A. Smith$^5$, 
Richard J. Terrile$^6$, Rodger I. Thompson$^2$}

\affil{$^1$University of California, Los Angeles, CA}

\affil{$^2$University of Arizona, Tucson, AZ}

\affil{$^3$Infrared Processing and Analysis Center (IPAC), Pasadena, CA}

\affil{$^4$University of Pennsylvania, Philadelphia, PA}

\affil{$^5$Institute for Astronomy (IFA), University of Hawaii, Honolulu, HI}

\affil{$^6$Jet Propulsion Laboratory (JPL), Pasadena, CA}

\begin{abstract}

We present the discovery of a candidate substellar object in a survey of 
young stars in the solar vicinity using the sensitivity and 
spatial resolution afforded by the NICMOS coronagraph on the Hubble 
Space Telescope. The H$=$12.1 mag object 
was discovered approximately 2$''$ from the TW Hydrae Association 
member CoD $-$33$^{\circ}$7795 (TWA 5), and the 
photometry implies a spectral type M8$-$M8.5, 
with a temperature of $\sim$2600K.
We estimate that the probability of 
a chance alignment with a background object
of this nature is $< 2 \times 10^{-5}$, and therefore
postulate the object (TWA 5B) is physically associated at 
a projected separation of 100 AU. Given the likely youth of the primary 
($\sim$ 10 Myr), current brown dwarf cooling models predict a mass 
of $\approx$ 20 Jupiter masses for TWA 5B.

\end{abstract}


\newpage

\section{Introduction}

The discovery and study of low mass objects in stellar systems is 
one of the key goals of contemporary observational astronomy. 
The substellar mass range from 10 to 80 Jupiter
masses (0.01 $-$ 0.08 M$_{\odot}$), 
of which no objects were known 10 years ago, is crucial to our
understanding of both stellar and planetary formation. Today, a handful of such 
objects are known. To make further progress, 
we are conducting an imaging 
survey of young, main-sequence stars to search for substellar 
companions using infrared coronagraphic techniques. Substellar objects cool with
age because they do not sustain hydrogen fusion and 
become more difficult to detect with time as they become both fainter and 
redder (c.f. Burrows et al. 1997). Using independently determined ages 
and distances for the target stars, the masses of detected secondaries 
can be ascertained from infrared magnitudes and 
theoretical evolutionary tracks on the H$-$R diagram. 

Over the last few years mounting evidence has suggested that a number
of young, active stars in the vicinity of TW Hydrae form a physical
association with an age of $\sim$ 10 Myr (Kastner et al. 1997; 
Webb et al. 1998a; Soderblom et al. 1998).
At an approximate distance of 50 pc, 
the ``TW Hydrae Association'' (TWA) is the region of 
recent star formation nearest to the Sun (Kastner et al 1997). 
Recently, Webb et al. (1998a)
added HR 4796 and identified five new systems (seven members), in which each 
system is characterized by the presence of X-ray emission, 
\ha emission, and strong lithium absorption associated with young stars.
The currently identified eleven
systems are shown to have similar space motions implying physical association and
a possible common origin (Webb et al. 1998b).
Webb et al (1998b) identify \cd, a member of this association, as a
likely spectroscopic binary, and although chromospheric activity can be enhanced in 
short-period, tidally-locked binaries (Barrado Y Navascues \& Stauffer 1996), 
they argue it is a member due to the presence of a strong Li line and consistent 
space motion with other TWA members. In their listing of the members of the TW Hya
Association, Webb et al. (1998a) use the provisional designation TWA 5
for \cd, which we also adopt. 

In this letter we present observations of TWA 5B, the 
first substellar object discovered with the coronagraph on the Near-Infrared Camera 
and Multi-Object Spectrometer (NICMOS) aboard the Hubble Space Telescope (HST). 
We discuss observation and reduction techniques which may prove important to other 
coronagraphic programs. We then present the measured and inferred properties 
from HST/NICMOS and Keck/LRIS \& NIRC observations of the newly discovered 
companion. TWA 5B was discovered independently at the IRTF using speckle imaging as 
reported in a companion paper by Webb et al (1998a).

\section{Observations}

\subsection{NICMOS}

TWA 5 (\cd; RA=11H 31M 55.3S, DEC=$-$34$\deg$ 36$'$ 27$"$ (J2000.0); M1.5V) 
was observed with NICMOS on 1998 April 25, 07:43 $-$ 08:40 UT. 
We imaged the star using the coronagraph in
Camera 2 (pixel scale  = $\sim$0.076 arcsec pixel$^{-1}$) in combination with a 
wide-band F160W 
filter (central wavelength: 1.59~$\mu$m, $\Delta\lambda$ = 0.40~$\mu$m), which 
corresponds closely to a Johnson H-band photometric filter. 
We obtained two multiple-exposure images with the bright primary star TWA 5 
behind the coronagraph (radius = 0.3~arc~sec) at two orientations differing
by 29.9 degrees. While the stellar point-spread-function (PSF), 
the instrumental scattering function, and detector artifacts 
rotate with the aperture, any real features in the unocculted area of the 
detector will be unaffected by a change in the camera orientation. Subtraction 
of these two images has been shown to further reduce residual PSF background
light (Schneider et al. 1998). Three standard NICMOS STEP64 MultiAccum (non-destructive read) 
integrations (MacKenty et al. 1997) totaling 684 s were executed at each orientation. 

The NICMOS coronagraphic images were reduced and processed
utilizing calibration darks and flat-fields created by the 
NICMOS Instrument Design Team (IDT) from on-orbit 
observations. The raw image data were calibrated in an analog 
to the pipeline reduction software, CALNICA (Bushouse 1997), which 
performs a dark subtraction, fits the multiaccum data using linear 
regression, and flat-fields the images. No
explicit background correction was made at this step because
the differencing of the two coronagraphic images removed 
scattered light and any background light. Two additional steps
were performed before flat-fielding; 
one process which used the first reads in the multiaccum sets
to estimate the signals in pixels which saturated in later reads,
and another which removed an estimate of the residual DC offsets
left after subtracting the dark current. Following initial calibration, bad
pixel values were replaced by a weighted interpolation with a radius of
5 pixels.  The three multiaccums at each orientation were then averaged 
to create final calibrated images. The images from each of the 
two spacecraft orientations were then aligned and subtracted from each other leaving 
lower amplitude residual noise near the coronagraphic hole edge, as well 
as positive and negative conjugates of any objects in the field of view. 
The raw data were reduced independently in a similar manner 
using the NICRED program (McLeod 1997).

\subsection{Keck}

We obtained an I-band ($\lambda = 0.73$ to $0.92\micron$) 
image of TWA 5 on 1998 February 6 UT on the Keck II 10m telescope 
using the Low Resolution Imaging Spectrometer (LRIS, Oke et al. 1995) 
in imaging mode. A short (1 sec), direct image was taken, but even 
in this minimal integration time, the primary star saturated, 
causing some bleeding. 
Under the assumption that the LRIS PSF is azimuthally
symmetric, the image was rotated by 180 degrees and subtracted from itself to search 
for possible bright, close secondaries.  

TWA 5 was observed on 18 January 1997 UT using the Near Infrared Camera, NIRC,
(Matthews \& Soifer 1994) on the Keck I 10m telescope. 
Two short exposure (0.43 sec), direct images
using both J ($\lambda = 1.1$ to $1.4\micron$) and K ($\lambda = 2.0$ to $2.4\micron$) 
filters were taken, bias subtracted and flat-field corrected. Sky subtraction 
was accomplished via the standard technique of 
dithering a few arcseconds and differencing the two frames.

\section{Results}

Subtraction and analysis of the NICMOS coronagraphic images reveal a 
stellar-like object (TWA 5B) at 
a separation of 1.96$\arcsec$ $\pm$ 0.01, and a PA of 1.79$\deg$ $\pm$ 0.36 
from TWA 5 (TWA 5A) (Figure 1a). This secondary is point-like with
a Full-Width-Half-Maximum (FWHM) of 0.$\arcsec$14, derived from the coronagraphic images, 
similar to the primary's FWHM of 0.$\arcsec$15, derived from the acquisition images. 
The secondary is also seen in the 
LRIS (Figure 1b) and NIRC (not shown) direct images at the same 
position angle and separation.
The positions of the primary and secondary in the NICMOS images 
were found by a least-squares isophotal
ellipse fitting process around the PSF core with a radius of 7 pixels to
exclude flux from any close objects. Since the target star is occulted in 
the NICMOS coronagraphic images, its position is ascertained from the target 
acquisition image, which resulted in independent measurements of each offset and 
position from both orientations.

\subsection{NICMOS}

In order to measure the flux of the secondary in the NICMOS images, 
the image conjugates (positive and negative) were separated
after subtraction, then shifted (translated, not rotated) with 
bi-cubic interpolation/resampling to null-out/align the PSF
of the companion.  The positive and negative images were combined to 
produce a reconstructed image of the companion with the background suppressed
except for residual photon noise (Figure 1a). 
 
The magnitude of TWA 5B was measured 
using a 15$\times$15 pixel square aperture centered on the companion. A correction
factor of 13.73$\%$, determined from coronagraphic photometric curves-of-growth
developed by the NICMOS IDT, was applied to the measured flux to compensate for
the flux which fell out of this aperture. With a conversion factor 
for the F160W filter of 2.20 $\times\ 10^{-6}$ Jy/ADU/s, and 
1087 Jy corresponding to an H magnitude of zero (Rieke 1998), the 
H magnitude of TWA 5B is 12.13 $\pm$ 0.05 mag. The photometry of 
the NICRED reduced images is consistent with these results
with a measured H magnitude of 12.16 $\pm$ 0.06 mag 
for TWA 5B. The uncertainty is dominated by NICMOS's calibration in relation
to standard stars. For the remainder of the paper we will use an
H magnitude of 12.14 $\pm$ 0.06 for TWA 5B (Table 1).

The H-band magnitude of TWA 5A was determined from aperture 
photometry of the two calibrated target acquisition images (at each 
of the two spacecraft orientations) processed as described in section 
2.1. Both measurements overlapped within the uncertainties and 
were averaged to yield H $=$ 7.2 $\pm$ 0.1 mag (Table 1).

\subsection{Keck}

The I-band flux of TWA 5B was measured from the one, direct LRIS image.
In the rotated and subtracted image (Figure 1b), 
a circular aperture with a radius 
of 3 pixels (0.$\arcsec$45) was used to minimize the
influence of spillover light from the primary. The flux measured in the aperture 
was corrected relative to an 
8 pixel (1.$\arcsec$2) radius aperture using bright objects in the LRIS field 
of view. Comparing the measured signal with a white dwarf, HZ4, whose I=14.7 mag 
(Zuckerman \& Becklin 1987), we find 
I= 15.8 $\pm$ 0.2 mag for TWA 5B. The majority of the uncertainty 
in the measure lies in the aperture 
correction due to the proximity of the saturated primary.

As in the LRIS images, the primary is also saturated in the NIRC direct images, 
but circular aperture photometry was 
performed on the secondary using small apertures to decrease the
influence of light from the primary. 
The measured J and K magnitudes of TWA 5B (Table 1)
harbor systematic uncertainties resulting from poorly calibrated 
curves of growth of both the object and standard stars. 
We estimate these uncertainties to be $\approx$ 0.2 mag.

\begin{center}
{\bf Table 1: Measured Photometry of TWA 5 A\&B}
\vspace{ .1in}

\begin{tabular}{lccccc}
\hline
\hline
Source &Spec type        & I  &  J & H & K  \\
\hline
 5A   &  M1.5  &  8.8 $\pm$0.4\tablenotemark{a}    
& 7.7$\pm$0.1\tablenotemark{a} & 7.2 $\pm$0.1 &  6.8$\pm$0.1\tablenotemark{a} \\
 5B   &$\sim$M8.5  & 15.8$\pm$0.2  &  12.6$\pm$0.2  & 12.14$\pm$0.06 & 11.4 $\pm$0.2 \\
\hline
\tablenotetext{a}{from Webb et al 1998}
\end{tabular}
\end{center}
$^a$from Webb et al. 1998a

\section{Discussion}

\subsection{Likelihood of Companionship}

With our data set it is not possible to prove a physical association of TWA 5A and TWA 5B. 
However, we can compare the apparent brightness (J=12.6 mag) and red color (I$-$K$=$4.4) with
objects found in various other infrared surveys.
From the Kirkpatrick et al. (1998a) survey of the solar neighborhood,
we find $\approx$ 0.2 stars per square arcminute at J $<$ 13 mag at the galactic latitude of 
TWA 5A (b$=$25) (we extrapolate from their completeness limit of J $<$ 17 mag using Lilly \& 
Cowie's (1987) galactic distribution model). Hence the a priori
probability of finding an object of this brightness within a 2$\arcsec$ 
radius circle (area = 0.0035 sq. arc
minutes) of any given point is about 0.06\%.
In an infrared survey of the Pleiades, Simons \& Becklin (1992) 
found colors for $\approx$ 500 background objects in 275 square degrees; 
only 3$\%$ had I$-$K $>$ 3.5 mag.
Most of those red objects were likely background galaxies which would be resolved by HST. 
Therefore, we estimate the probability that a red background star is separated from
TWA 5A by 2 arcsec is the product of these two, or $<$ 2 $\times\ 10^{-5}$.

We consider the possibility that TWA 5B could be a foreground, low-mass star. 
Assuming a spectral type later than
M7 from the colors, we find M$_{H}$ $>$ 10.15 for a dwarf star (Kirkpatrick \& McCarthy 1994), 
so the photometric distance would be 25 parsecs.  Henry (1991), 
in a volume limited infrared survey, finds 6 objects with M$_{H}$ $>$ 9.5 within 
5 parsecs from the sun. Assuming a 
spherical distribution of low mass stars in the solar neighborhood, 
we should expect 750 such objects out to 25 parsecs, so the a priori probability
of finding one in projection within a 2$\arcsec$ radius circle is 2 $\times\ 10^{-8}$. 

Given these small probabilities, it is unlikely we would find a foreground or 
background object of this nature in our sample size of less than 50 stars. 
The combination of a small separation from another low-mass star, an extremely red color, and 
moderate brightness is rare at a galactic latitude of 25 degrees. Further, as shown below, 
TWA 5B falls on the same isochrone as TWA 5A on the H-R diagram, 
implying coevality, and therefore companionship. 
For the remainder of the paper, we assume TWA 5B is physically associated with TWA 5A.

\subsection{Effective Temperature and Bolometric Luminosity}

The photometric measurements place strong constraints on the nature of the
secondary.  The colors of TWA 5B are consistent with a spectral type of M8-M8.5.  
(Kirkpatrick \& McCarthy 1994; Luhman et al. 1997).  
An effective temperature is required to position TWA 5B on an H$-$R diagram, 
but the temperature scale for late, young M-dwarfs is uncertain (Allard et al. 1997). 
Kirkpatrick et al (1993) match synthetic spectral fits to observed 
spectra for late type stars and derive a temperature of 2875K and 2625K 
for M8 and M9, respectively. 
Luhman \& Rieke (1998) extrapolate from Leggett et al's (1996) model fits to derive 2505K   
for this spectral class, which agrees with the newer models used by Leggett et al (1998).
Simons et al (1996) present a derived relation between color and effective temperature which
gives 2538K for the J$-$K color we measure. Since many young, 
low-mass stars possess spectral features indicative of both dwarfs 
and giants (Luhman \& Rieke 1998), it has 
been argued that adopting an intermediate effective temperature for 
the two luminosity classes, 100$\deg$K to 150$\deg$K hotter
than the dwarf scale, is appropriate (Luhman et al 1997; White et al 1998).
Perrin et al (1998) have found an M8 giant to have an effective 
temperature of 2806K from their model fits.  
With the uncertainty in temperature for late M dwarf stars and the added
uncertainty from the youth, we plot the range (Figure 2) appropriate to 
the colors we measure with a horizontal dotted line from 2505K to 2875K, which includes
the derived giant temperature. Assuming
coevality with TWA 5A, the primary's isochrone intersects this range at 
approximately 2600K.

There is no parallactic distance measured to TWA 5, so 
we adopt the approximate distance of 50 pc to the association.  
This is based on Hipparcos parallax measures to TW Hydra, 
HD98800, HR4796 and TWA 9, thought to be members of the same association. 
With an H-magnitude of 12.14 and a distance modulus of 3.49, we derive the
luminosity of TWA 5B to be 0.0021 $\pm$ 0.0003 L${\odot}$. In doing so, we adopt a 
bolometric correction of 2.8 $\pm$ 0.1 mag for an M8.5 dwarf, 
as suggested by Kirkpatrick et al. (1993).

\subsection{Derived Mass}

We place TWA 5A \& B on pre-main sequence 
evolutionary tracks (D$'$Antona \& Mazitelli 1997) to 
infer masses (Figure 2). The range of temperatures plotted 
places the mass of the secondary in the range
of 0.015 to 0.055 M${\odot}$ (15 to 55 \mj) with the primary's isochrone, 
assuming coevality,
intersecting at $\approx$ 20 \mj. Evolutionary tracks 
themselves do differ somewhat due to different models and 
atmospheres used. The tracks of (Baraffe et al. 1998) 
place the mass of the secondary in the range 15 to 60 \mj, 
with the primary's isochrone intersecting near 25 \mj. 
Burrows et al's (1997) models predict a 10 Myr old, 20 \mj\ brown dwarf 
will have an effective temperature
of 2609K and a luminosity of 0.0022 L${\odot}$, which is in 
good agreement with our data. 

\section{Implications for Formation of Brown Dwarf Companions}

We present spatially resolved, high signal-to-noise optical and 
near-infrared photometry 
of the young ($\sim$ 10Myr) TW Hydrae Association member
\cd (= TWA 5) with evidence for a approximately
20 \mj\ brown dwarf companion at a 
projected separation of 100 AU from the primary. 
If indeed the secondary is a true companion, it will be interesting to see if a 
young brown dwarf exhibits Li absorption as do all known members of 
the TWA (Webb et al. 1998a). It has been conjectured that 
brown dwarfs with mass $10-80$ \mj\ are rare at separations 
less than 5 AU 
(Marcy et al. 1998) and greater than 200 AU (Oppenheimer et al. 1999), 
but there are now at
least four likely substellar companions known at separations from 30 to 200 AU 
(Kirkpatrick et al. 1998b, 
Nakajima et al. 1995, White et al. 1998, this paper). Although orbiting diverse primaries 
(white dwarf, field M star, T Tauri star, young M star) and subject to small number 
statistics, these results hint that the semi-major axis distribution of brown dwarf 
and late M dwarf secondaries may differ 
since the latter are found throughout a wide range of semi-major axes.

This work is supported in part by NASA grants NAG 5-4688 to UCLA and 
NAG 5-3042 to the University of Arizona NICMOS Instrument Design Team.  
This Letter is based on observations obtained with the NASA/ESA Hubble Space Telescope 
at the Space Telescope Science Institute, which is operated by the
Association of Universities for Research in Astronomy, Inc. under NASA contract
NAS 5-26555. Some of the data presented herein were obtained at the W.M.Keck 
Observatory. We would like to thank R. White, A. Weinberger, and D. Weintraub for 
their valuable comments and the anonymous referee for comments 
which clarified the presentation.

\clearpage

\figcaption{(a) The NICMOS H-band image of TWA 5B (left). Observations at two 
different orientations have been subtracted with TWA 5A behind the coronagraph. The inset shows 
positive and negative images of TWA 5B rotated 29.9$\deg$ about the primary's center. 
Combining the two conjugate images of the secondary by image translation, the 
diffraction limited profile of the companion is seen. The residual light not 
rejected by the coronagraph appears in two positions from the translation centered on the 
hole locations. Pixels inside the hole have been set to gray. 
The orientation of the image corresponds to the primary on the 
left, marked with a star.
(b) An LRIS image (right) of TWA 5A which  
has been rotated by 180$\deg$ and subtracted from itself. 
The companion is seen to the north and its conjugate negative to the south (inset). 
There is residual light from incomplete subtraction
of the diffraction spikes to the east of the secondary. The
primary is saturated (see inset); the saturated pixels have been set to gray. 
\label{fig1}}

\figcaption[fig2a.eps]{Evolutionary tracks (D'Antona \& Mazzitelli 1997) 
with TWA 5A (open diamond) TWA 5B (squares) plotted. 
The dashed-horizontal line represents an uncertainty in assigning a photometric temperature 
to a young star (see text for discussion). The assumption
of coevality between TWA5A and TWA5B would have an isochrone intersect at a 
temperature $\approx$ 2600K and at a mass of 20 \mj.\label{fig2} }

{
\centering \leavevmode
\epsfxsize=.65\columnwidth \epsfbox{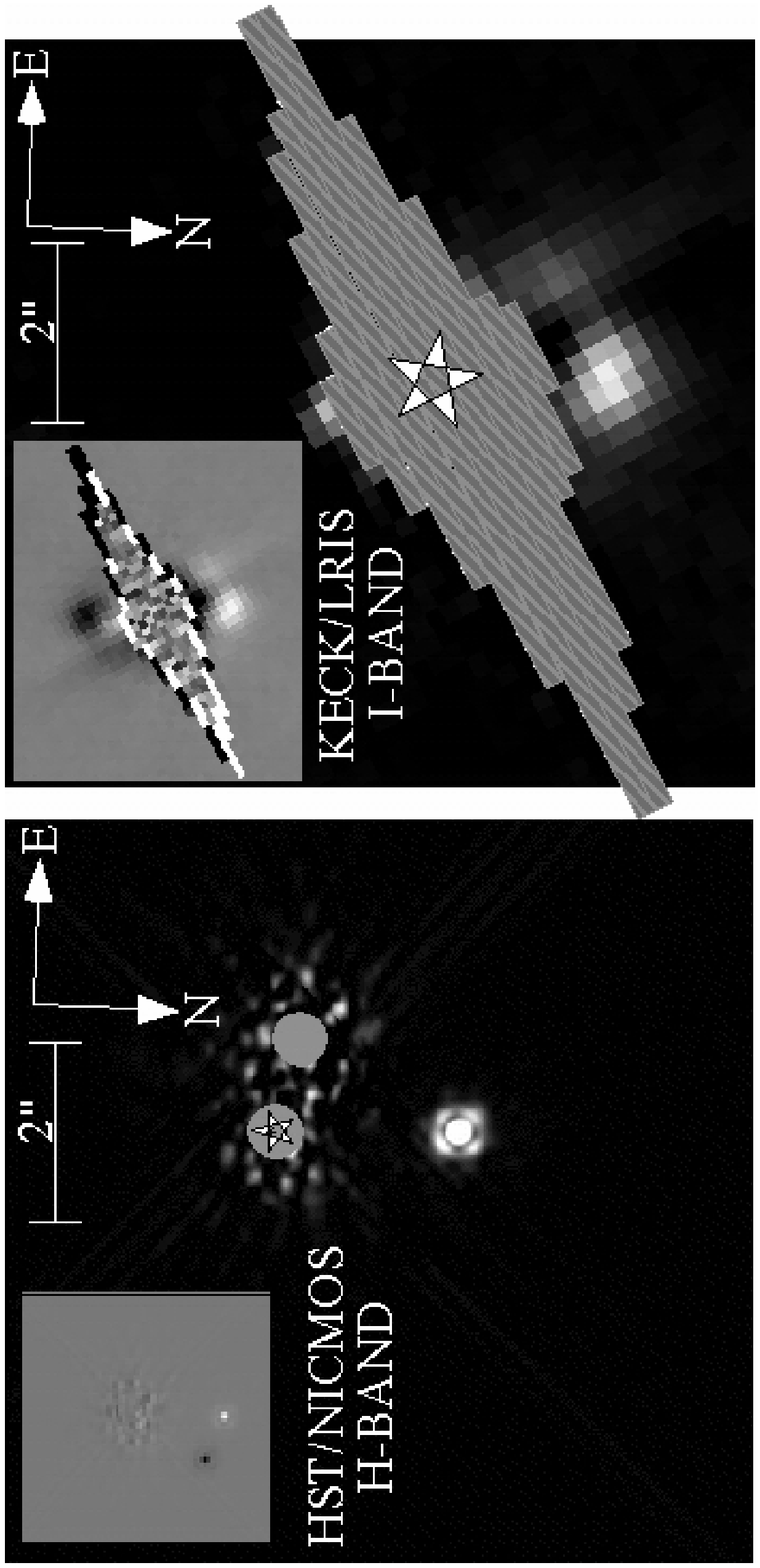}
}

{
\centering \leavevmode
\epsfxsize=.97\columnwidth \epsfbox{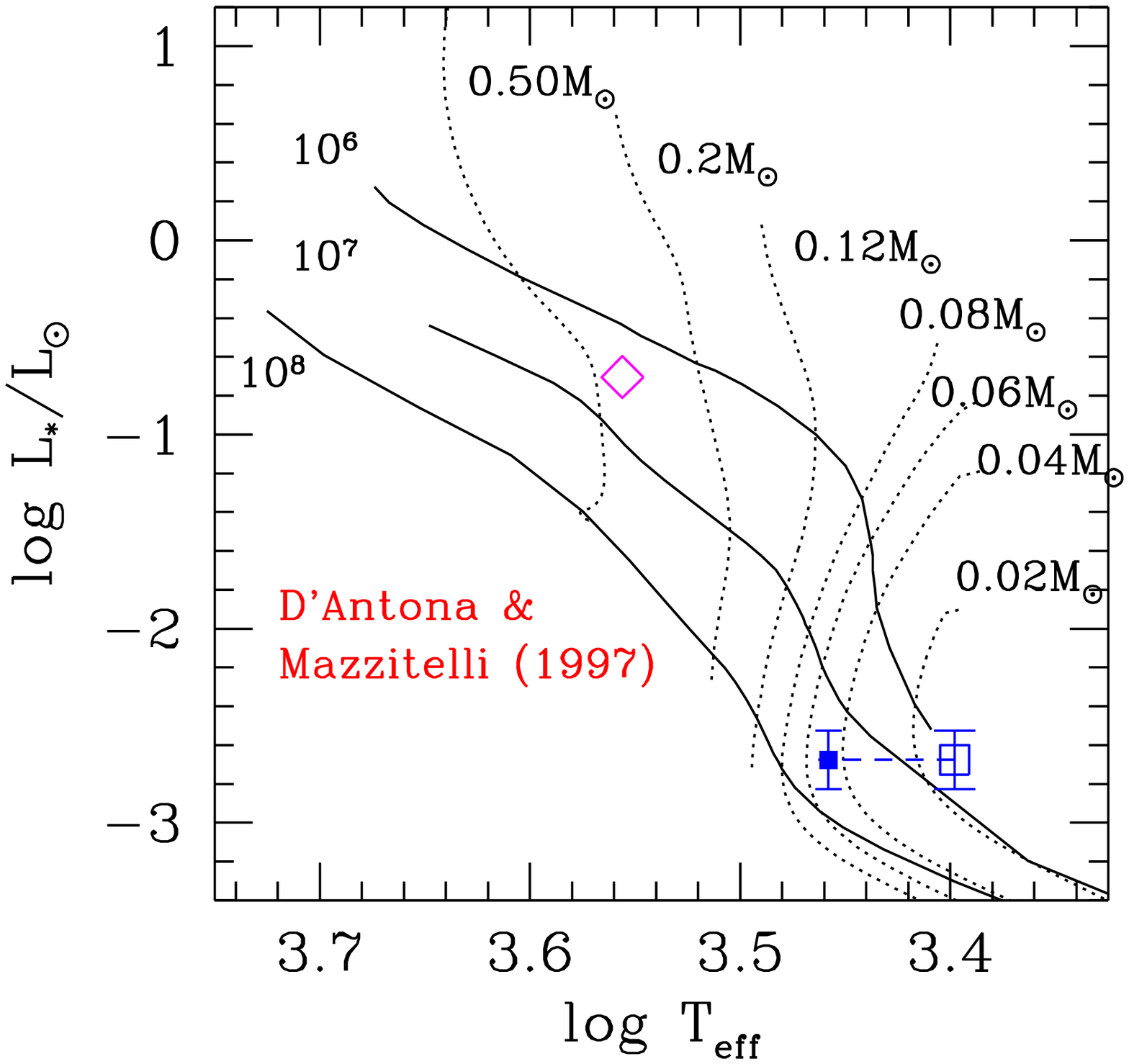}
}

\end{document}